\documentclass{article}
\usepackage{spconf,amsmath,graphicx}

\usepackage{enumitem}
\setlist{nosep, leftmargin=14pt}

\usepackage{color}
\usepackage{color, colortbl}
\usepackage{multirow}
\usepackage{makecell}
\usepackage{hyperref}

\definecolor{TableGray}{gray}{0.9}
\definecolor{TableLightCyan}{rgb}{0.88,1,1}

\usepackage{mwe} 

\title{On Enhancing Brain Tumor Segmentation Across Diverse Populations with Convolutional Neural Networks}
\name{Fadillah Adamsyah Maani, Anees Ur Rehman Hashmi, Numan Saeed, and Mohammad Yaqub}
\address{Mohamed bin Zayed University of Artificial Intelligence, Abu Dhabi, UAE}
\begin{document}
%
\maketitle
\begin{abstract}
Brain tumor segmentation is a fundamental step in assessing a patient's cancer progression. However, manual segmentation demands significant expert time to identify tumors in 3D multimodal brain MRI scans accurately. This reliance on manual segmentation makes the process prone to intra- and inter-observer variability. This work proposes a brain tumor segmentation method as part of the \textit{BraTS-GoAT} challenge. The task is to segment tumors in brain MRI scans automatically from various populations, such as adults, pediatrics, and underserved sub-Saharan Africa. We employ a recent CNN architecture for medical image segmentation, namely MedNeXt, as our baseline, and we implement extensive model ensembling and postprocessing for inference. Our experiments show that our method performs well on the unseen validation set with an average DSC of 85.54\% and HD95 of 27.88. The code is available on \url{https://github.com/BioMedIA-MBZUAI/BraTS2024_BioMedIAMBZ}.
\end{abstract}
\begin{keywords}
BraTS-GoAT, Brain MRI, Tumor Segmentation, MedNeXt, Model Ensembling
\end{keywords}

\section{Introduction}

Cancer is a leading cause of death worldwide, specifically brain tumors. Many initiatives have been presented to enhance automatic brain tumor segmentation, aiming to increase the quality of diagnosis. The Brain Tumor Segmentation (BraTS) challenge is one of the most notable initiatives in this area. However, previous BraTS challenges have been tailored to address specific types of tumors within a single patient demographic, e.g., Adult Glioma \cite{6975210,Bakas2017,DBLP:journals/corr/abs-2107-02314,bakaslgg,bakasgbm}, Pediatrics \cite{kazerooni2024brain}, sub-Saharan African brain glioma patient population \cite{adewole2023brain}, Brain Meningioma \cite{labella2023asnrmiccai}, and Brain Metastasis \cite{moawad2023brain}. To fill this gap, the organizer introduces a new challenge segment, namely BraTS Generalizability Across Tumors (BraTS-GoAT). This manuscript is our contribution to the BraTS-GoAT challenge. In summary, we implement the MedNeXt \cite{mednext} architecture to segment tumors, an ensembling mechanism to enhance prediction, and a set of postprocessing to remove prediction noises. We submitted our models for the testing phase using MedPerf \cite{Karargyris2023}.

\section{Methods}
\subsection{Dataset}

The dataset for this challenge was compiled from diverse populations, including adults, pediatrics, and underrepresented groups from sub-Saharan Africa. It comprises 2,251 brain MRI scans in the training set and 360 scans in the validation set. The provided MRI modalities are T1, T1Gd, T2, and T2-FLAIR. Each scan includes expert annotations that identify three tumor subtypes: enhancing tumor (ET), tumor core (TC), and whole tumor (WT). It is interesting to note that each MRI scan contains one or more tumors. The size of each MRI scan is consistently 240x240x155.

\subsection{Network Architecture}

\begin{figure*}[t!]
    \centering
    {\includegraphics[width=0.9\linewidth]{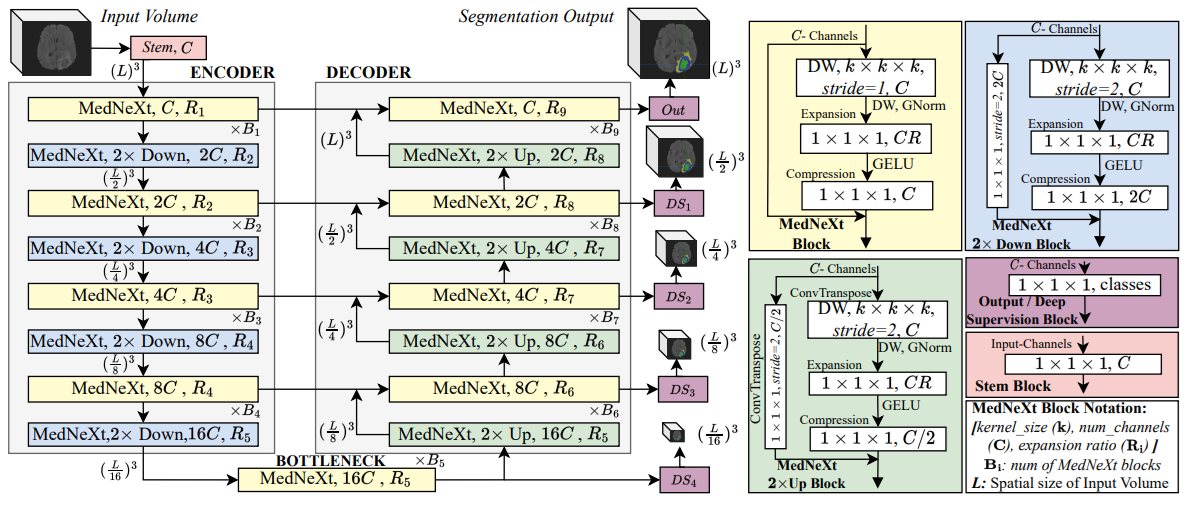}}
    \caption{The MedNeXt~\cite{mednext} architecture. It combines the benefits of CNNs and transformers by designing transformer-inspired ConvNeXt \cite{convnext} blocks for image segmentation tasks.
    }
    \label{fig:mednext}
\end{figure*}

We implemented MedNeXt, the successor to the renowned nnU-Net framework, as our baseline architecture. MedNeXt draws inspiration from the ConvNeXt \cite{convnext} architecture, which is prominent in 2D natural image classification. Major components of MedNeXt include residual connections, depth-wise convolution with a variety of large kernel sizes ($k \in \{3,5,7,9\}$), and point-wise convolution with wide channels. MedNeXt offers a range of model sizes: small (S), base (B), medium (M), and large (L). Further details on this architecture can be found in \cite{mednext}. For this competition, the model input channel is 4, and the output channel is 3, corresponding to TC, WT, and ET.

\subsection{Inference}
Given that the typical size of a brain MRI scan exceeds the model's input sizes, we employed a sliding window inference approach to segment tumor(s). Additionally, to enhance the reliability of predictions, we implemented test-time augmentation (TTA) by flipping the scans across all possible dimensions ($2^3=8$) and averaging the predicted tumor probabilities. Further improvement in segmentation accuracy was achieved through model ensembling, leveraging models trained using 5-fold cross-validation (CV), and aggregating their probability outputs. We devised a set of post-processing steps to refine our predictions and reduce the false positives (FPs) in tumor detection. For each output channel (i.e. TC, WT, and ET), we first conducted a connected component analysis to cluster predicted voxels into groups (i.e. tumors), estimated the size of these groups, and computed the average probability of tumor voxels within each group. Subsequently, we eliminated groups characterized by small sizes and low mean tumor voxel probabilities.

\section{Experimental Setup}
\subsection{Model Training}
We conducted our experiments on an NVIDIA GPU with 24 GB of memory. The models' input size is 128x128x128 voxels. Employing 5-fold cross-validation on the training set, we developed MedNeXt-B and MedNeXt-M models. Both models were initially trained with a kernel size of 3 for 55 epochs, after which kernel upsampling to 5 was applied, and the models were further trained for an additional 90 epochs. The optimization was carried out using the AdamW optimizer, set with a learning rate of 3e-4 and a weight decay of 1e-6. We employed a cosine annealing scheduler, initiating with a warm-up period of 3 epochs, a start learning rate of 1e-7 during the warm-up, and concluding with a final learning rate of 1e-6. We used a batch size of 2 for training. In terms of the objective function (loss), deep supervision, brain MRI preprocessing, and data augmentation, we followed \cite{maani2024advanced}.

We tuned postprocessing hyperparameters using Wandb \cite{wandb} sweep on the 5-fold CV and then manually tuned them to adapt to the unseen validation set.

\subsection{Model Inference}
We begin by preprocessing the input MRI scans: first, by cropping them to isolate the brain region and eliminate background, and then by normalizing the intensity of each MRI channel using mean and standard deviation, excluding background pixels in this process. Tumor probability maps are generated through sliding window inference, employing a 50\% overlap between windows to enhance the precision of our predictions. Subsequently, we improve the robustness of these predictions by averaging the tumor probability masks from all models in an ensemble approach, resulting in a refined final tumor probability map. We then postprocess the tumor probability map to detect and segment tumors.

{
\begin{table*}[h!]
\centering
\caption{
Summary of our results on the validation leaderboard. Each row represents a submission with potentially different sets of post-processing hyperparameters. \textit{Full training} refers to models trained using all training samples (with 5 different seeds) instead of 5-fold CV.
}
\label{tab:val}
\resizebox{\textwidth}{!}{
\begin{tabular}{l|rrrr|rrrr|rrrr|rrrr}
\rowcolor{TableGray} & \multicolumn{8}{|c}{\textbf{BraTS 2024 Score}} & \multicolumn{8}{|c}{\textbf{Legacy Score}} \\
\rowcolor{TableGray} & \multicolumn{4}{|c|}{Dice} & \multicolumn{4}{c|}{HD95} & \multicolumn{4}{c}{Dice} & \multicolumn{4}{|c}{HD95} \\
\rowcolor{TableGray} \multirow{-3}{*}{} & \multicolumn{1}{|c}{ET} & \multicolumn{1}{c}{TC} & \multicolumn{1}{c}{WT} & \multicolumn{1}{c|}{Avg} & \multicolumn{1}{c}{ET} & \multicolumn{1}{c}{TC} & \multicolumn{1}{c}{WT} & \multicolumn{1}{c}{Avg} & \multicolumn{1}{|c}{ET} & \multicolumn{1}{c}{TC} & \multicolumn{1}{c}{WT} & \multicolumn{1}{c}{Avg} & \multicolumn{1}{|c}{ET} & \multicolumn{1}{c}{TC} & \multicolumn{1}{c}{WT} & \multicolumn{1}{c}{Avg} \\
\hline
\makecell[l]{MedNeXt-B5} & 0.7755 & 0.8380 & 0.8335 & 0.8157 & 53.24 & 30.91 & 40.63 & 41.59 & 0.8306 & 0.8699 & 0.9026 & 0.8677 & 24.83 & 14.98 & 12.36 & 17.39 \\
\makecell[l]{\,\,\,\,+ postprocessing} & 0.8292 & 0.8450 & 0.8682 & 0.8475
& 33.77 & 31.29 & 26.39 & 30.48 & 0.8450 & 0.8671 & 0.9004 & 0.8708 & 24.75 & 21.04 & 13.44 & 19.74 \\
\makecell[l]{MedNeXt-M5} & 0.8381 & 0.8485 & 0.8716 & 0.8527 & 30.94 & 29.53 & 25.35 & 28.61 & 0.8509 & 0.8697 & 0.9041 & 0.8749 & 23.88 & 19.21 & 11.81 & 18.30 \\
\makecell[l]{\,\,\,\,+ Sweep for ET} & 0.8390 & 0.8485 & 0.8716 & 0.8530 & 30.45 & 29.53 & 25.35 & 28.44 & 0.8511 & 0.8697 & 0.9041 & 0.8750 & 23.79 & 19.21 & 11.81 & 18.27 \\
\makecell[l]{\,\,\,\,+ Full training} & 0.8397 & 0.8501 & \bfseries 0.8746 & 0.8548 & 30.69 & 29.31 & \bfseries 24.30 & 28.10 & 0.8524 & 0.8713 & 0.9063 & 0.8767 & 25.62 & 20.97 & 13.58 & 20.06\\
\makecell[l]{\,\,\,\,+ Post. hyp. tuning} & \bfseries 0.8405 & \bfseries 0.8510 & \bfseries 0.8746 & \bfseries 0.8554 & \bfseries 30.36 & \bfseries 28.97 & \bfseries 24.30 & \bfseries 27.88 & 0.8524 & 0.8713 & 0.9063 & 0.8767 & 24.89 & 20.24 & 12.85 & 19.33 \\
\end{tabular}
}
\end{table*}

\begin{figure*}[h!]
    \centering
    \includegraphics[width=0.9\linewidth]{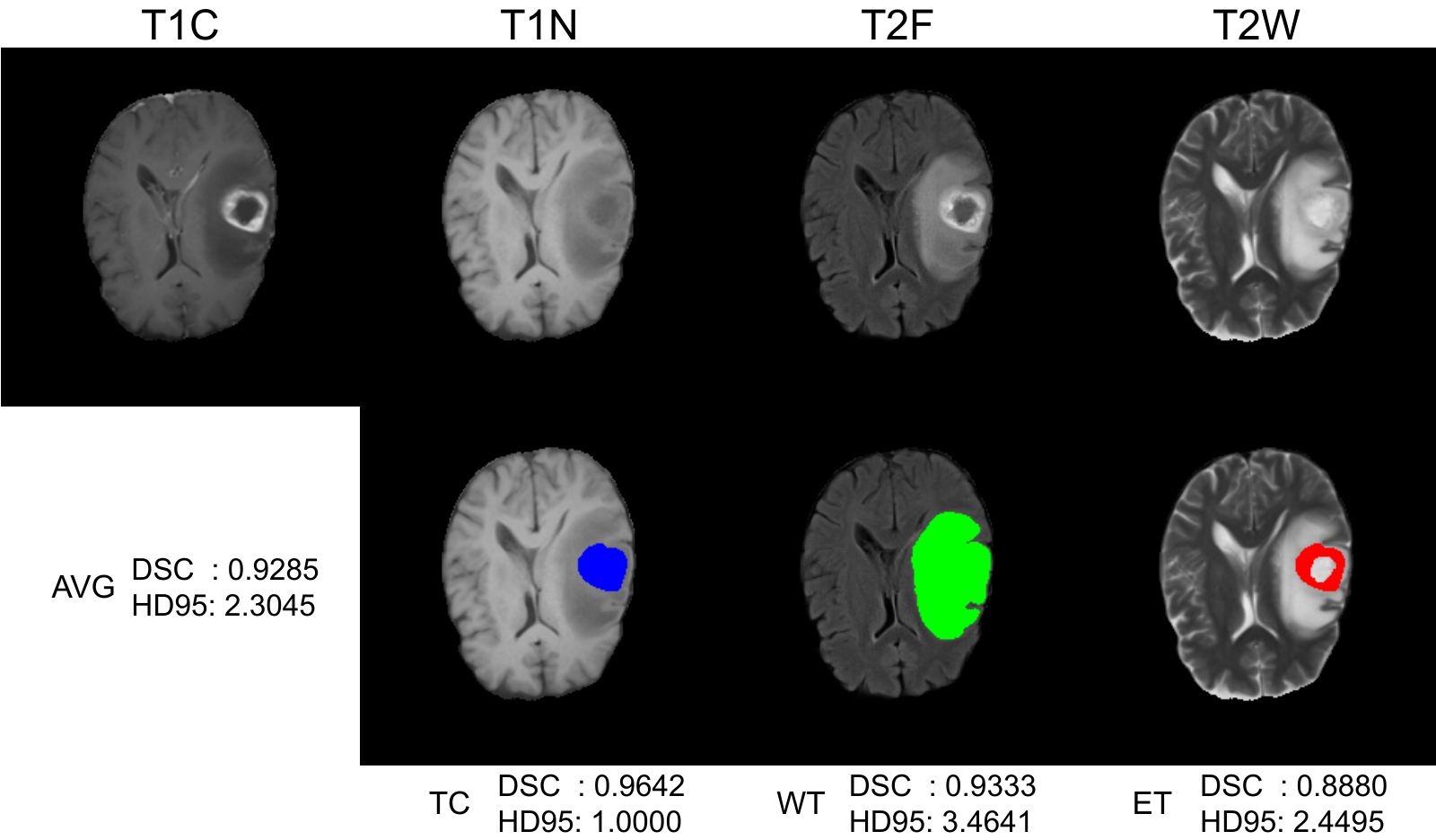}
    \caption{Qualitative result showing median performance on the validation leaderboard.}
    \label{fig:qual}
\end{figure*}
}

\section{Results and Discussion}



We summarize our results on the validation leaderboard in Table \ref{tab:val}. We provide a qualitative result for the median case of our submission in Figure \ref{fig:qual}. We began by utilizing MedNeXt-B with a kernel size 5, trained using 5-fold cross-validation (CV), and averaged the tumor probability maps from these models' predictions. Our post-processing step enhances the average Dice Similarity Coefficient (DSC) by 3.18\%. Employing a larger model, MedNeXt-M, further boosts the average DSC by 0.52\%. Based on our experiments, automatic hyperparameter tuning with Weights and Biases (Wandb) sweeps did not significantly improve performance on the validation leaderboard. We hypothesize that the sweeps are less effective because they tune the hyperparameters based solely on the 5-fold CV results without ensembling, whereas ensembling is implemented when submitting to the validation leaderboard.

Training with the full dataset enhanced model performance, particularly for the WT class. In this study, we trained our models using five different seeds. We applied a 0.5 threshold to each channel (ET, TC, and WT). For postprocessing, we removed detected tumors with sizes smaller than 100 voxels for ET, 150 voxels for TC, and 500 voxels for WT.

\section{Conclusion}
This work represents our contribution to the BraTS-GoAT competition. We utilized a CNN-based model to detect tumors from brain MRI scans. The model processes four MRI input channels and produces three output channels for TC, WT, and ET, respectively. We implemented model ensembling and postprocessing techniques to enhance predictions and reduce noise. Our experiments demonstrated that larger models perform better in this competition, suggesting that the BraTS-GoAT competition is more challenging than previous BraTS competitions.

\section{Compliance with Ethical Standards}
This work adhered to the BraTS2024-GoAT data policies to prevent any potential ethical misconduct.

\section{Acknowledgments}

We would like to thank Ikboljon Sobirov and Mariam Aljuboory for contributing to our group in the BraTS 2023 competition.


\bibliographystyle{IEEEbib}
\bibliography{refs}
\end{document}